\documentclass{article}
\usepackage{graphicx, caption, amsmath}
  \usepackage{setspace}
\usepackage[margin=1in]{geometry}
\usepackage{authblk}
\DeclareCaptionLabelFormat{nospace}{#1#2}

\begin{document}
\title{GPA-Tree: Statistical Approach for Functional-Annotation-Tree-Guided Prioritization of GWAS Results}
\author[1]{\small Aastha Khatiwada}
\author[1]{\small Bethany J. Wolf}
\author[2]{\small Ayse Selen Yilmaz}
\author[1, 3]{\small Paula S. Ramos}
\author[2]{\small Maciej Pietrzak}
\author[1]{\small Andrew Lawson}
\author[1]{\small Kelly J. Hunt}
\author[4]{\small Hang J. Kim}
\author[2, *]{\small Dongjun Chung}

\affil[1]{\footnotesize Department of Public Health Sciences, Medical University of South Carolina, Charleston, South Carolina, USA}
\affil[2]{\footnotesize Department of Biomedical Informatics, The Ohio State University, Columbus, Ohio, USA}
\affil[3]{\footnotesize Department of Medicine, Medical University of South Carolina, Charleston, South Carolina, USA}
\affil[4]{\footnotesize Division of Statistics and Data Science, University of Cincinnati, Cincinnati, Ohio, USA}
\affil[*]{\footnotesize To whom correspondence should be addressed (chung.911@osu.edu).}

\vspace{2ex}

\date{}

\onehalfspacing
\normalsize
\maketitle
\abstract{\textbf{Motivation:} 
In spite of great success of genome-wide association studies (GWAS), multiple challenges still remain. 
First, complex traits are often associated with many single nucleotide polymorphisms (SNPs), each with small or moderate effect sizes. 
Second, our understanding of the functional mechanisms through which genetic variants are associated with complex traits is still limited. 
functional annotations related to risk-associated SNPs.
To address these challenges, we propose GPA-Tree and it simultaneously implements association mapping and identifies key combinations of functional annotations related to risk-associated SNPs by combining a decision tree algorithm with a hierarchical modeling framework.\\
\textbf{Results:} First, we implemented simulation studies to evaluate the proposed GPA-Tree method and compared its performance with existing statistical approaches. The results indicate that GPA-Tree outperforms existing statistical approaches in detecting risk-associated SNPs and identifying the true combinations of functional annotations with high accuracy. Second, we applied GPA-Tree to a systemic lupus erythematosus (SLE) GWAS and functional annotation data including GenoSkyline and GenoSkylinePlus. The results from GPA-Tree highlight the dysregulation of blood immune cells, including but not limited to primary B, memory helper T, regulatory T, neutrophils and CD8$^+$ memory T cells in SLE. These results demonstrate that GPA-Tree can be a powerful tool that improves association mapping while facilitating understanding of the underlying genetic architecture of complex traits and potential mechanisms linking risk-associated SNPs with complex traits.\\
\textbf{Availability:} The GPATree software is available at https://dongjunchung.github.io/GPATree/.\\
}

\section{Introduction}
As of February $2021$, the genome-wide association studies (GWAS) Catalog has published $4,865$ GWAS studies and identified $247,051$ genotype-trait associations (https://www.ebi.ac.uk/gwas/) \cite{buniello2019nhgri}. While the number of identified genotype-trait associations has increased substantially in the last decade, there are multiple challenges that still need to be addressed to improve identification of genotype-trait associations and to prioritize the likely causal variants. 
First, it has been shown that a complex trait can be influenced by multiple single nucleotide polymorphisms (SNPs), often with small or moderate effect sizes \cite{price2015progress, nikpay2015comprehensive}. Such SNPs often do not meet the genome-wide \textit{p-value} cutoff of $5 \times 10^{-8}$ and as a result, still many trait-associated SNPs remain unidentified. In theory, utilizing a large sample size will improve statistical power to detect these SNPs. However, traits of limited prevalence in the population often result in limited sample sizes, and recruiting a large sample size requires significant resources and is often not  feasible. Therefore, there is a critical need to find alternate ways to increase statistical power to detect SNPs with small and moderate effect sizes.

Second, over $85\%$ of the genetic variants identified by GWAS are located in non-coding regions \cite{giral2018into} and it is often difficult to understand their functional roles in the trait etiology. For example, in autoimmune diseases, about $90\%$ of the causal genetic variants lie in non-coding regions, a bulk of which are located in regulatory DNA regions \cite{maurano2012systematic, farh2015genetic}. Utilizing tissue- and cell type-specific functional information can potentially improve our understanding of biological mechanisms through which SNPs may be associated with traits \cite{Schork2013a}. The general hypothesis is that functional roles relevant to trait-associated SNPs may influence the distribution of these SNPs in the GWAS summary statistics. Therefore, integrating GWAS summary statistics and functional annotation information might not only improve statistical power to detect SNPs 
, but also identify the mechanisms by which trait-associated SNPs may influence trait \cite{Ming2018, Zablocki2014, Schork2013a}. For example, in the case of autoimmune diseases like systemic lupus erythematosus (SLE) and multiple sclerosis (MS), risk-associated SNPs might be more enriched for those with roles in immunity, while for psychiatric disorders like bipolar disorder (BPD) and schizophrenia (SCZ), risk-associated SNPs might be more relevant to the central nervous system or brain function. 

Recognizing the potential of such integrative approaches, several methods have been proposed to prioritize SNPs and identify relevant functional annotations by integrating GWAS data with functional annotation data \cite{Schork2013a, Zablocki2014, lu2015statistical, GenoWAP2015, Ming2018, Ming439133}. Still currently available methods utilize functional annotations in a relatively simple form without considering interactions between them, i.e., only main effects terms are included in the model. However, this can be a critical limitation because valuable in-depth biological insight can often be obtained through investigating combinations of the functional annotations, e.g., different types of histone modifications, epigenetic marks in different immune cell subsets, and expression quantitative trait loci (eQTL) for different traits. In theory, some existing methods can be extended by including interaction terms to identify combinations of functional annotations. However, this requires strong prior scientific knowledge, which is often lacking, especially when a large number of functional annotations is considered in the analysis. Moreover, including all possible interactions can quickly become computationally taxing. Therefore, there is a critical need for a method that can efficiently identify relevant combinations of functional annotations without requiring strong prior knowledge. 

To fill in this important research gap, we propose GPA-Tree, a novel statistical approach that simultaneously prioritizes trait-associated SNPs and identifies key combinations of functional annotations related to the mechanisms through which trait-associated SNPs influence the trait, within a unified framework. Specifically, GPA-Tree is based on a hierarchical modeling approach integrated with a decision tree algorithm and facilitates easy interpretation of findings. GPA-Tree takes GWAS summary statistics as input, which allows wide applications and adaptations. Our comprehensive simulation studies and real data applications show that GPA-Tree consistently improves statistical power to detect trait-associated SNPs and also effectively identifies biologically important combinations of functional annotations.

\section{GPA-Tree}
\subsection{Model}\label{gpatreemodel}

Let $\mathbf{Y}_{M \times 1} = (Y_1, Y_2, \ldots, Y_M)'$ be a vector of genotype-trait association \textit{p}-values for $i = 1, 2,\cdots, M$ such that $y_i$ denotes the \textit{p}-value for the association of the $i^{th}$ SNP with the trait. We also assume that we have K binary annotations ($\mathbf{A}$).
$$
\mathbf{A} = ({\mathbf{A}}_{.1}, \ldots ,  {\mathbf{A}}_{.K}) = \begin{pmatrix} 
a_{11} & \hdots & a_{1K} \\
\vdots & \ddots& \vdots \\
a_{M1} & \hdots & a_{MK} 
\end{pmatrix}_{M \times K} \: \text{, \: where}
$$
 \begin{equation*}
  a_{ik} =
    \begin{cases}
      0, & \text{if $i^{th}$ SNP is not annotated in the $k^{th}$ annotation}\\
      1, & \text{if $i^{th}$ SNP is annotated in the $k^{th}$ annotation}
    \end{cases}       
\end{equation*}

Here our ultimate goal is association mapping, i.e., identifying SNPs associated with the trait given both GWAS and functional annotation data. To accomplish this, we introduce the latent variable $\mathbf{Z}$, where $z_i$ indicates association of $i^{th}$ SNP with the trait. Then, the GWAS association \textit{p}-values ($y_i$) are assumed to come from a mixture of non-risk-associated ($z_i = 0$) and risk-associated groups ($z_i = 1$). As previously proposed by Chung and colleagues \cite{chung2014gpa}, if the $i^{th}$ SNP belongs to the non-risk-associated group ($z_i=0$), then its \textit{p}-value is assumed to come from the Uniform distribution on $[0, \: 1]$. This is based on the rationale that $U[0,\: 1]$ provides  a \textit{p}-value density corresponding to the non-risk-associated group \cite{pounds2003estimating}. If the $i^{th}$ SNP belongs to the risk-associated group ($z_i=1$), then its \textit{p}-value is assumed to come from the Beta distribution with parameters ($\alpha,\: 1$), where $0 < \alpha < 1$. We restrict $\alpha$ in the Beta distribution to be between 0 and 1 because the smaller $\alpha$ value corresponds to the higher density at lower \textit{p}-values, while the $\alpha$ value closer to one resembles a $Unif[0,\: 1]$ distribution.

We further integrate functional annotation data with the GWAS data by modeling the latent $\mathbf{Z}$ as a function of the functional annotation data $\bf{A}$. Specifically, we define a function $f$ that is a combination of functional annotations $\mathbf{A}$ and relate it to the expectation of latent $\mathbf{Z}$ as given in Equation (\ref{eq:1}).
\begin{equation}
    P(Z_i=1 ; a_{i1}, ..., a_{iK}) = \text{\it{f}}(a_{i1}, ..., a_{iK})
    \label{eq:1}
\end{equation}
Let $\theta = (\alpha, \boldsymbol{\pi})$, where $\boldsymbol{\pi} = \{\pi_1, \pi_2, ..., \pi_M\}$ is a function of $\mathbf{A}$ and represents the prior probabilities that the SNPs belong to the risk-associated group, i.e., $\pi_i = P(Z_i = 1)$. See Section 1 in the Supplementary Materials for the joint distribution of the observed data, and the incomplete and complete data log-likelihoods.

\subsection{Algorithm}

Given the approach described in Section \ref{gpatreemodel}, we implemented parameter estimation using an EM algorithm. The function $f$ in Equation (\ref{eq:1}) is estimated by a decision tree algorithm and it allows to identify combinations of functional annotations related to risk-associated SNPs. To improve stability, we employed a two-stage approach for parameter estimation. Specifically, in Stage 1, we first estimate the parameter $\alpha$ without identifying a combination of functional annotations. Then, in Stage 2, we identify key combinations of functional annotations ($f(\mathbf{A})$) while the parameter $\alpha$ is kept fixed as the value obtained in the first step. We illustrate more detailed calculation steps below.\\

\hspace{-0.4cm}\textbf{Stage 1: }\\
For the $i^{th}$ SNP, the $t^{th}$ iteration of the E-step can be written as:
\begin{equation}
\label{eq:2}
\begin{array}{l@{}l}
\bf{E-step}:  
	z_i^{(t)} = E [Z_i; \: \mathbf{Y}, \mathbf{A}, {\boldsymbol{\theta}}^{(t-1)}] \\ 
	\hspace{2.1cm}	= Pr( Z_{i}=1; \: \mathbf{Y}, \mathbf{A}, \boldsymbol{\theta}^{(t-1)})\\
	\hspace{2.1cm}=\frac{P(Y_i;\: Z_i=1, \:  \boldsymbol{\theta}^{(t-1)}) P(Z_i = 1;  \, \mathbf{A}_{i.},  \: \boldsymbol{\theta}^{(t-1)}) } {\sum\limits_{d \in{\{1, 0\}}} P(Y_i; \:Z_i=d, \: \boldsymbol{\theta}^{(t-1)}) P(Z_i = d;  \: \mathbf{A}_{i.},\: \boldsymbol{\theta}^{(t-1)})} \\
	\hspace{2.1cm}	= \frac{\alpha^{(t-1)}y_{i}^{\alpha^{(t-1)}-1} \pi_{i}^{(t-1)}}{1-{\pi_{i}}^{(t-1)} + \alpha^{(t-1)}y_{i}^{\alpha^{(t-1)}-1} \pi_{i}^{(t-1)} } 
		\end{array}
\end{equation}
In the $t^{th}$ iteration of the M-step, ${\pi_i}$ and $\alpha$ are updated as:
\begin{equation}
\label{eq:3}
\begin{array}{l@{}l}
\bf{M-step}: \text{Fit a linear regression model as} \\
\hspace{1.85cm} {z}_i^{(t)} = \beta_{0}^{(t)} + \beta_{1}^{(t)} {a}_{i1} + \cdots + \beta_K^{(t)} {a}_{iK} + \epsilon_i^{(t)} \\
\hspace{1.55cm} \text{Update } {\pi}_i^{(t)}  \text{as the predicted value from the linear regression model.} \\
\hspace{1.55cm} \text{Update } \alpha^{(t)}= - \sum\limits_{i=1}^{M} {z_{i}}^{(t)} / \sum\limits_{i=1}^{M} {z_{i}}^{(t)} log(y_{i}) \text{,} \\
\end{array}
\end{equation} 
where $\beta_k^{(t)}, k = 0, \cdots, K$ are the regression coefficients and $\epsilon_i^{(t)}$ is the error term. 
The E and M steps are repeated until both the incomplete log-likelihood and the $\alpha$ estimate converge. The $\alpha$ and $\boldsymbol{\pi}$ estimated in this stage are used to fix $\alpha$ and initialize $\boldsymbol{\pi}$, respectively, in Stage 2.\\

\hspace{-0.4cm}\textbf{Stage 2:}\\
In this stage, we implement another EM algorithm employing a decision tree algorithm (CART \cite{breiman1984classification}), which allows to identify union, intersection, and complement relationships between functional annotations in estimating $\pi_i$. 

For the $i^{th}$ SNP, the $t^{th}$ iteration of the E-step can be written as:
\begin{equation}
\label{eq:4}
\begin{array}{l@{}l}
{\bf{E-step}}:  z_i^{(t)} = \frac{\hat{\alpha}y_{i}^{\hat{\alpha}-1} \pi_{i}^{(t-1)}}{1-{\pi_{i}}^{(t-1)} + \hat{\alpha}y_{i}^{\hat{\alpha}-1} \pi_{i}^{(t-1)} } 
		\end{array}
\end{equation}
Note that here $\alpha$ is fixed as $\hat{\alpha}$, which is the final estimate of $\alpha$ obtained from Stage 1.
In the $t^{th}$ iteration of the M-step, ${\pi_i}$ is updated as:
\begin{equation}
\label{eq:5}
\begin{array}{l@{}l}
\bf{M-step}: \text{Fit a CART model as} \\
\hspace{1.85cm} {z}_i^{(t)} = f^{(t)}(a_{i1}, \cdots, a_{iK}) + \epsilon_i^{(t)} \\
\hspace{1.55cm} \text{Update } {\pi}_i^{(t)}  \text{as the predicted value from the CART model,} \\
\end{array}
\end{equation}
where $\epsilon_i$ is the error term. In the M-step, the complexity parameter ($cp$) is the key tuning parameter and defined as the minimum improvement that is required at each node of the tree. Specifically, in the CART model, the largest possible tree (i.e., a full-sized tree) is first constructed and then pruned using $cp$. The pruned regression tree structure identified by the CART model upon convergence of the EM algorithm (Equation (\ref{eq:5})) is used as $f$ in Equation (\ref{eq:1}). This approach allows for the construction of the accurate yet interpretable regression tree that can explain relationships between functional annotations and genotype-trait associations.  The E and M steps are repeated until the incomplete log-likelihood converges. 

We note that unlike the standard EM algorithm, the incomplete log-likelihood in Stage 2 is not guaranteed to be monotonically increasing. Therefore, we implement Stage 2 as a generalized EM algorithm by retaining only the iterations in which the incomplete log-likelihood increases compared to the previous iteration.

\subsection{Prioritization of risk-associated SNPs  and identification of relevant combinations of functional annotations}
Once the parameters are estimated as described in Section 2.2, we can now prioritize risk-associated SNPs and identify combinations of functional annotations relevant to these SNPs. First, SNPs are prioritized using the local false discovery rate, $fdr$, which is defined as the posterior probability that the $i^{th}$ SNP belongs to the non-risk-associated group given its GWAS \textit{p}-value and functional annotation information, i.e., $ fdr(Y_{i}, \mathbf{A}_{i.}) =  P(Z_{i} = 0; \:Y_{i}, \mathbf{A}_{i.}) = 1 - P(Z_{i} = 1;\:Y_{i}, \mathbf{A}_{i.})$. We utilize the `direct posterior probability' approach \cite{newton2004detecting} to control the global false discovery rate, $FDR$, which is the expected ratio of the number of SNPs that are incorrectly predicted to be risk-associated SNPs (false positives) compared to the number of SNPs that are predicted to be risk-associated SNPs (positives). In this approach, SNPs are first sorted by their $fdr$ in an ascending order, denoted as $h_i$. The threshold for $fdr$, $\kappa$, is then increased from 0 to 1 until 
\begin{equation*}
\label{eq:6}
FDR = \frac{\sum\limits_{i=1}^M h_i  \, 1\{h_i \leq \kappa\}}{\sum\limits_{i=1}^M 1 \{h_i \leq \kappa\}} \leq \tau ,
\end{equation*}
where $\tau$ is the predetermined level of $FDR$ (e.g., $\tau \leq 0.05)$. Finally, SNPs with $h_i \leq \kappa $ are considered to be risk-associated SNPs. Second, relevant combinations of annotations are inferred based on the combination of functional annotations selected by the CART model upon convergence of the EM algorithm in Stage 2.  

\subsection{ShinyGPATree: Shiny app for interactive analysis of risk-associated SNPs and the functional annotation tree}
\begin{figure}[ht]
\centering \includegraphics[width=\linewidth]{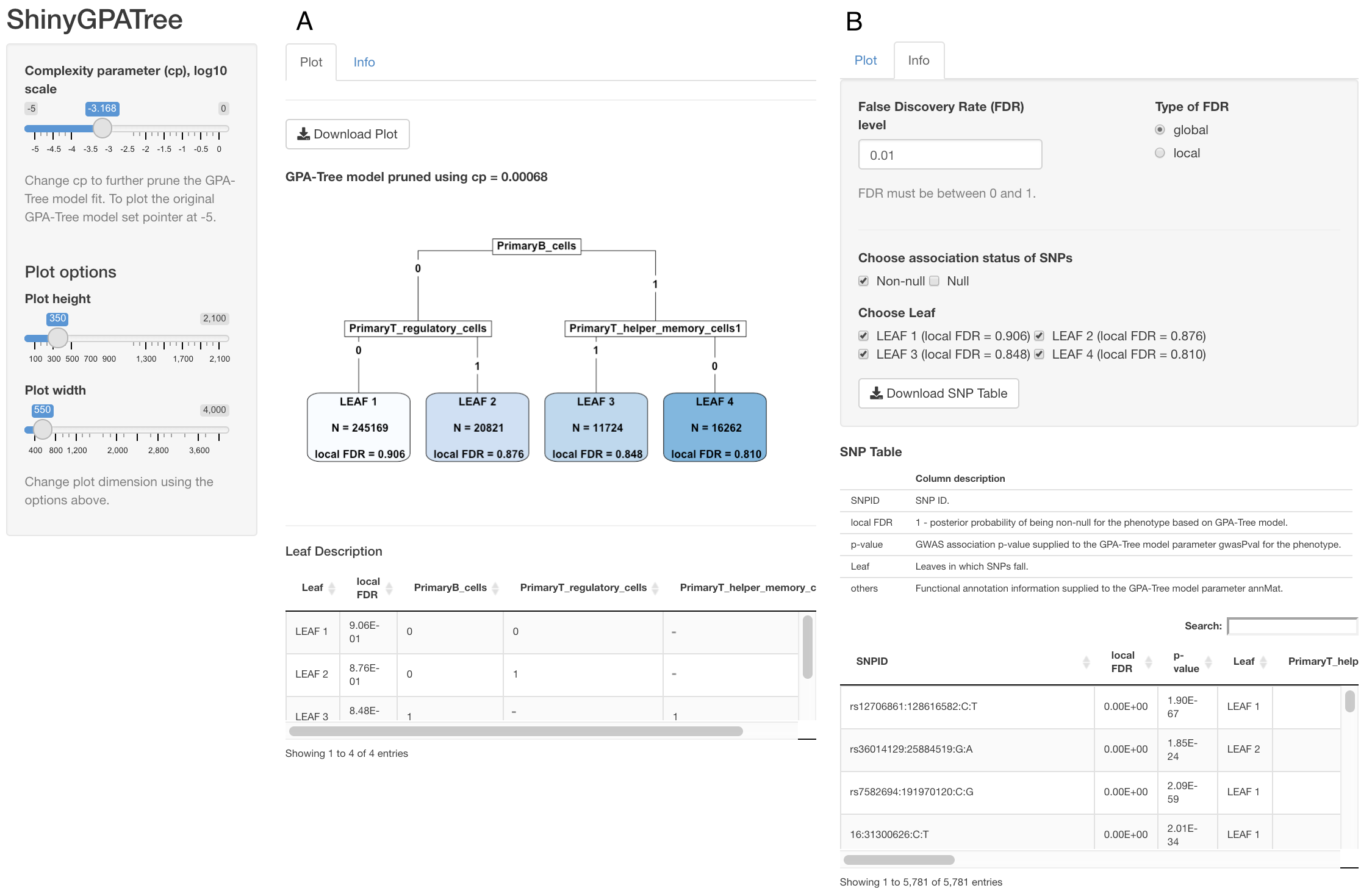}
\caption{Screenshot of the ShinyGPATree app with (A) the `Plot' tab and (B) the `Info' tab open.}
 \label{fig:Figure2}
\end{figure}
We implemented the forementioned GPA-Tree algorithm as an R package \texttt{‘GPATree’}. To further facilitate user’s convenience, we developed \texttt{`ShinyGPATree’}, a Shiny app for interactive analysis of risk-associated SNPs and the functional annotation tree (Fig~\ref{fig:Figure2}). This Shiny app can be open by sequentially running \texttt{GPATree()} and \texttt{ShinyGPATree()} functions. First, the \texttt{GPATree()} function takes 4 arguments: \texttt{gwasPval, annMat, initAlpha and cpTry}. \texttt{gwasPval} is a $M \times 1$ matrix of GWAS association \textit{p}-values for $M$ SNPs, \texttt{annMat} is a $M \times K$ matrix of $K$ binary functional annotations for $M$ SNPs, \texttt{initAlpha} is the initial alpha value to be used to fit the GPA-Tree model (default value = 0.1), and \texttt{cpTry} is the $cp$ parameter to be used to fit the GPA-Tree model (default value = 0.001). The \texttt{GPATree()} function generates a GPA-Tree model fit required for the ShinyGPATree app. The \texttt{ShinyGPATree()} function takes the output of \texttt{GPATree()} as an input and opens the ShinyGPATree app using the R code below.\\
{\small{\ttfamily R> fit <- GPATree(gwasPval, annMat, initAlpha, cpTry)\\
R> ShinyGPATree(fit)}}\\
The ShinyGPATree app provides visualization of the GPA-Tree model fit, identifies risk-associated SNPs, and characterizes the combinations of functional annotations that can describe the risk-associated SNPs. The app also allows to improve the visualization of the GPA-Tree model fit by collating or separating layers of the model using the $cp$ parameter. The number of non-risk-associated and risk-associated SNPs that can be characterized by combinations of functional annotations are also automatically updated based on user-selected $cp$, FDR type (global vs. local) and FDR level values. The interactive nature of the app allows users to effortlessly interact with the GPA-Tree model results to generate plots, prioritize risk SNPs, and make inferences about relevant combinations of functional annotations for the risk-associated SNPs. ShinyGPATree consists of two main tabs, namely `Plot’ and `Info’, which are explained in detail below.

\subsubsection{Plot tab: Visualization of the GPA-Tree model}

Fig~\ref{fig:Figure2}A shows the layout of the ShinyGPATree app, where the `Plot' tab opens by default. In the displayed plot, each leaf (terminal node) is characterized by combinations of the functional annotations that are encountered as users move from the root node to the leaf. The summary information is provided for each leaf, including the number of SNPs that satisfy the combination of functional annotations specific to the leaf and the mean local FDR for these SNPs. The summary information displayed in each leaf is automatically updated as the user modifies the \texttt{cp} value on the left panel. Users can also improve visualization of the functional annotation tree plot using the \texttt{Plot width} and \texttt{Plot height} options on the left panel. The \texttt{`Download Plot'} button on the top allows users to download the functional annotation tree plot as a PNG format file. Finally, a table titled `Leaf Description' underneath the plot characterizes the functional annotations that are $0$ or $1$ for SNPs in specific leaves. 

\subsubsection{Info tab: Association mapping and annotation selection}
The `Info' tab opens the user interface for association mapping and functional annotation characterization for SNPs as seen in Fig~\ref{fig:Figure2}B. Under this tab, users can find more information on specific SNPs driving the visualization. The top of the panel provides multiple options to control association mapping, including FDR level and FDR type (global vs. local). It also provides options to select which SNPs to display, e.g., choosing SNPs that fall on specific leaves of the GPA-Tree model and/or selecting SNPs with specific association status (non-risk-associated vs. risk-associated SNPs). The `SNP Table' at the bottom of the `Info' tab panel shows information about the SNPs that satisfy these options. Each row of the table represents a SNP, where columns include SNP ID, local FDR value, GWAS association \textit{p}-value, the leaf ID in which the SNP is located, and the corresponding complete functional annotation information. The \texttt{`Download SNP Table'} button allows users to download the `SNP Table' as a CSV format file.

\begin{figure*}[ht]
\centerline{\includegraphics[width=\linewidth]{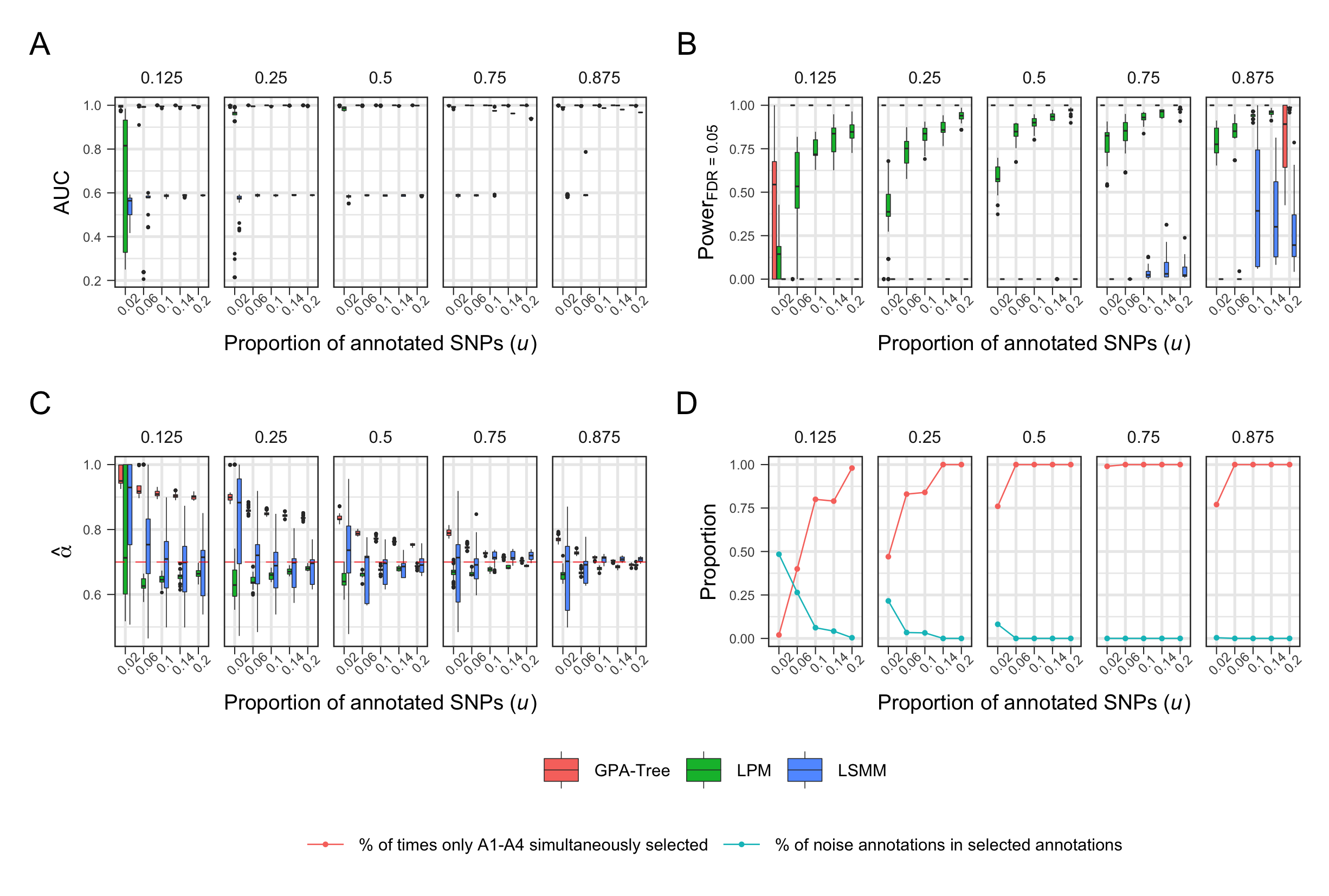}}
\caption{Comparison of (A) AUC, (B) statistical power to detect true risk-associated SNPs when global FDR is controlled at the nominal level of $0.05$, (C) estimated $\alpha $ parameter, and (D) proportion of times only true functional annotations $A_1-A_4$ are simultaneously identified by GPA-Tree (red line) and the average proportion of noise annotations ($A_5-A_{75}$) among the functional annotations identified by GPA-Tree (blue line). The results are presented for different proportions of SNPs annotated in $A_1-A_4$ ($u$; x-axis) and proportions of the overlap between SNPs annotated in $A_1 - A_2$ and $A_3 - A_4$ ($v$; panel). $M = 100,000$, $K=75$, and $\alpha =0.7$ in $Beta(\alpha, 1)$ and results are summarized from 100 replications.}
\label{fig:Figure 4}
\end{figure*}

\section{Results}
\subsection{Simulation study}
We conducted a simulation study to evaluate the performance of the proposed GPA-Tree approach. For all the simulation data, the number of SNPs was set to $M = 100,000$, the number of annotations was set to $K = 75$, and risk-associated SNPs were assumed to be characterized with the combinations of functional annotations defined by $L = (A_1 \cap A_2) \cup (A_3 \cap A_4)$; all the remaining functional annotations ($A_k, k = 5, .., 75$) were considered to be noise annotations. The percentage of annotated SNPs ($u$) for annotations $A_1 - A_4$ was set to $2\%, 6\%, 10\%, 14\%$ and $20\%$, while the percentage of overlap between the true combinations of functional annotations ($v$) was set to $12.5\%, 25\%, 50\%, 75\%$ and $87.5\%$. For noise annotations $A_5 - A_{75}$, approximately $20\%$ of SNPs were annotated by first generating the proportion of annotated SNPs from $Unif[0.1, 0.3]$ and then randomly setting this proportion of SNPs to one. The SNPs that satisfy the functional annotation combination $L$ were assumed to be risk-associated SNPs and their \textit{p}-values were simulated from $Beta(\alpha, \: 1)$ with $\alpha = 0.7$. The SNPs that do not satisfy $L$ were assumed to be non-risk SNPs and their \textit{p}-values were simulated from $U[0, \: 1]$.

For each combination of the simulation parameters defined above, we simulated $100$ datasets and compared the performance of GPA-Tree with LPM \cite{Ming439133} and LSMM \cite{Ming2018}.  The metrics for comparing the methods include (1) area under the curve (AUC), where the curve was created by plotting the true positive rate (sensitivity) against the false positive rate (1-specificity) to detect risk-associated SNPs when global FDR was controlled at various levels; (2) statistical power to identify risk-associated SNPs when global FDR was controlled at the nominal level of 0.05; and (3) estimation accuracy for $\alpha$ parameter in the $Beta(\alpha, \: 1)$ distribution used to generate the \textit{p}-values of risk-associated group.  For GPA-Tree, we also examined the accuracy of detecting the correct functional annotation tree, based on the proportion of simulation data for which all relevant functional annotations in L ($A_1-A_4$) were identified simultaneously, and the average proportion of noise functional annotations ($A_5-A_{75}$) among the functional annotations identified by GPA-Tree. Here we especially investigate how the percentage of SNPs annotated in $A_1-A_4$ ($u$) and the overlap between SNPs annotated in $A_1 - A_2$ and $A_3 - A_4$ ($v$) impact GPA-Tree's ability to separate functional annotations relevant to the risk-associated SNPs from noise annotations.  

\begin{itemize}
    \item {\bf AUC: } Fig \ref{fig:Figure 4}A shows the AUC comparison between GPA-Tree, LPM, and LSMM. For all the combinations of $u$ and $v$, GPA-Tree showed the consistently higher AUC relative to LSMM while performing comparably or better than LPM. The performance of LPM and LSMM improved as signal-to-noise ratio increases (i.e., as $u$ and $v$ increase), demonstrating performance closer to GPA-Tree.
    
    \item {\bf Statistical power: } Fig \ref{fig:Figure 4}B compares the power to detect true risk-associated SNPs when global FDR is controlled at $0.05$ for the three methods. GPA-Tree showed higher statistical power to detect true risk-associated SNPs relative to LPM and LSMM for almost all combinations of $u$ and $v$. The estimated power for GPA-Tree was relatively more variable for $u=2\%$ and $v=12.5\%$ but it still outperformed LPM and LSMM. The statistical power of LPM increased as a function of $u$ for all $v$, and the statistical power of LSMM increased as $u$ increases for higher $v$. However, both LPM and LSMM showed greater variability in statistical power compared to GPA-Tree and on average they showed lower statistical power compared to GPA-Tree.
    
    \item {\bf Estimation of parameter $\boldsymbol{\alpha}$:} Fig \ref{fig:Figure 4}C shows the $\alpha$ parameter estimates obtained from the three methods. GPA-Tree showed less variability in the $\alpha$ estimates compared to LPM and LSMM.  LPM was on average more accurate than GPA-Tree in estimating $\alpha$, however it still often underestimated $\alpha$. LSMM showed decreased variability in estimation of $\alpha$ as $u$ increases, and estimated $\alpha$ well for higher $u$ and $v$ levels. GPA-Tree generally overestimated $\alpha$ and this was most notable when $u$ and $v$ are small. As $u$ and $v$ increase, $\alpha$ estimates from GPA-Tree became closer to the true value. When $u$ and $v$ are large ($u \geq 10\%$ and $v \geq 75\%$), GPA-Tree estimated $\alpha$ accurately. We note that overestimation of $\alpha$ by GPA-Tree did not impact the method’s ability to identify the true combinations of functional annotations or the risk-associated SNPs, which are the main objectives of GPA-Tree.
    
    \item {\bf Selection of relevant and noise annotations:} The red line in Fig \ref{fig:Figure 4}D shows the proportion of times only functional annotations in the true combination L ($A_1 - A_4$) were simultaneously identified by GPA-Tree while the blue line shows the proportion of noise annotations ($A_5 - A_{75}$) that were also selected. Excluding instances when signal in the data is really weak ($u \leq 6\%$ and $v \leq 25\%$), GPA-Tree successfully identified all functional annotations included in the true combination $L$ more than $75\%$ of the time. Moreover, GPA-Tree could identify all functional annotations included in the true combination approximately $100\%$ of the time as $u$ or $v$ get larger (Fig \ref{fig:Figure 4}D, red line). These results demonstrate the potential of GPA-Tree to correctly identify true annotations as long as signal in the data is not too weak. In instances where GPA-Tree did not identify all functional annotations included in $L$, it either identified one or more noise annotations in addition to the true annotations (false positives), or failed to identify one or more annotations in $L$ (false negative) (Fig \ref{fig:Figure 4}D, blue line).\vspace*{1pt}

\end{itemize}

\subsection{Real data analysis}

We applied the GPA-Tree approach to the SLE GWAS data \cite{langefeld2017transancestral} sourced from the GWAS Catalog \cite{buniello2019nhgri} (https://www.ebi.ac.uk/gwas/). Summary statistics were originally obtained using the genotyped and imputed Immunochip, profiled for $18,264$ individuals ($6,748$ cases and $11,516$ controls) of European ancestry. $336,745$ SNPs passed quality control criteria. After excluding SNPs located in the MHC region, $293,976$ SNPs were included in the final analysis and integrated with functional annotation data from GenoSkyline (GS) \cite{lu2016integrative} and GenoSkylinePlus (GSP) \cite{lu2017systematic}. GS were generated by integrating epigenetic annotations from the Roadmap Epigenomics Consortium \cite{kundaje2015integrative}. They predict tissue-specific functional relevance for SNPs, which are available for seven tissue clusters (brain, gastrointestinal/GI, lung, heart, blood, muscle and epithelium tissues). GSP added another layer of information to GS in the form of epigenomic and transcriptomic annotations, and are available for $127$ annotation tracks. The Manhattan plot and \textit{p}-value histogram for SLE GWAS data are presented in Fig S1 in the Supplementary Materials. 

\begin{figure}[ht]
\centerline{\includegraphics[width=\linewidth]{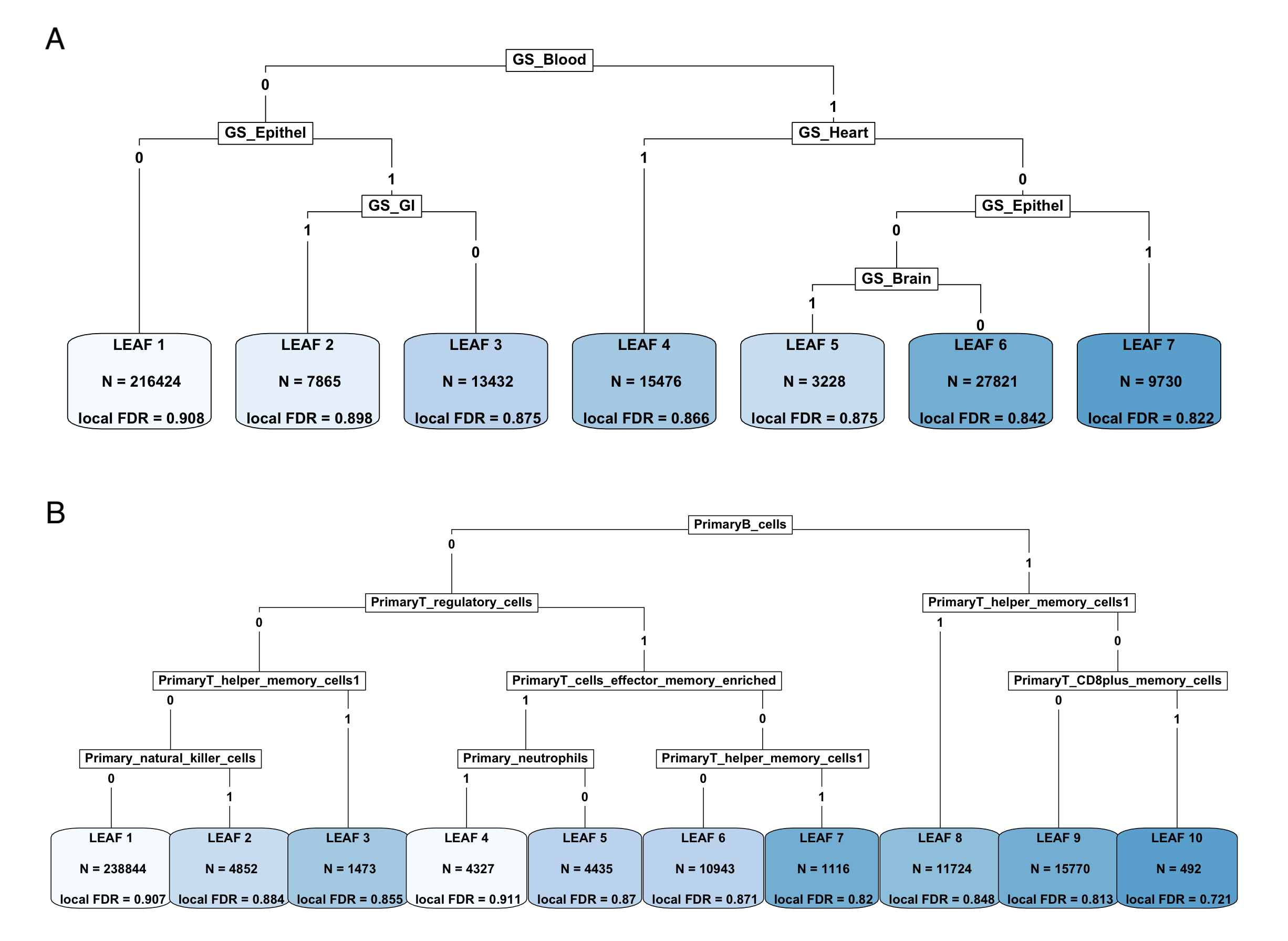}}
\caption{Functional annotation tree identified by GPA-Tree approach when (A) seven tissue-level GenoSkyline (GS) annotations and (B) 10 blood-related cell-type-level GenoSkylinePlus (GSP) annotations are considered. Both trees were generated by pruning the GPA-Tree model fit using $cp = 2.5 \times 10^{-4}$. Each leaf (terminal node) in the tree shows the total number of SNPs in the leaf and the mean local FDR for the SNPs in the leaf.}
\label{fig:Figure 5}
\end{figure}

\subsubsection{Tissue-level investigation}
 We initially investigated the functional potential of all SNPs using seven tissue-specific GS annotations. With a GS score cutoff of $0.5$, $35.90\%$ of SNPs were annotated in at least one of the seven tissue types (Fig S2A in the Supplementary Materials) and the percentage of annotated SNPs ranged from $8.66\%$ for lung tissue to $19.14\%$ for blood tissue (Fig S2B in the Supplementary Materials). We also measured the overlap in SNPs annotated in different tissue types using log odds ratio (Fig S2C in the Supplementary Materials). While the highest proportion of SNPs is annotated for blood tissue, SNPs annotated for blood tissue overlap less with other tissue types. On the contrary, SNPs annotated for heart, lung and muscle tissues overlap more with other tissue types. This is consistent with the literature indicating that blood shows the lowest levels of eQTL sharing with other tissue types while muscle and lung tissues show higher levels of eQTL sharing \cite{gtex2015genotype, lu2016integrative}. 
Next, we applied GPA-Tree to the SLE GWAS and GS annotation data for association mapping and characterization of relevant functional annotations. GPA-Tree identified $8,962$ SLE-associated SNPs at the nominal global FDR level of $0.05$. Among SLE-associated SNPs, $46.40\%$ were annotated for at least one of the seven GS tissue type (Fig S3A in the Supplementary Materials), and the percentage of annotated SNPs ranged from $9.89\%$ for lung tissue to $30.22\%$ for blood tissue (Fig S3B in the Supplementary Materials). We also measured relative enrichment (RE), the ratio of the proportion of SLE-associated SNPs annotated for a specific tissue type, relative to the proportion of non-SLE-associated SNPs annotated for the same tissue type. RE was again highest for the blood tissue with the value of $1.61$ (Fig S3C in the Supplementary Materials). These results are consistent with the dysregulation of blood immune cells that characterizes SLE and other autoimmune diseases like Crohn’s disease, ulcerative colitis and rheumatoid arthritis \cite{lu2016integrative}. 

The original GPA-Tree model fit contained blood tissue at the root node and included 28 leaves. For easier interpretation, we used ShinyGPATree app to prune the tree so that it includes 7 leaf nodes (Fig \ref{fig:Figure 5}A). We note that although it is occasionally possible to obtain a large functional annotation tree that can be cumbersome to visualize and interpret, the ShinyGPATree app can be utilized to manage such cases as it allows users to investigate different layers of functional annotation trees in an interactive and dynamic manner. For example, the annotation combination for SNPs in leaf 7 can be written as Blood $\cap$ !Heart $\cap$ Epithelium, i.e., leaf 7 includes SNPs that are annotated for blood and epithelium tissues but not for heart tissue. The number of SNPs that are located in each leaf node, and the combination of functional annotations that describe SNPs in each leaf node are displayed in Fig \ref{fig:Figure 5}A. Further investigation of the GPA-Tree model fitting results showed that, among the $8,962$ SLE-associated SNPs, $578$ are concurrently annotated for blood and epithelium tissues while not being annotated for heart tissue as represented in leaf 7; $609$ are concurrently annotated for both blood and heart tissues as represented in leaf 4; and 230 are concurrently annotated for epithelium and GI tissues while not being annotated for blood tissue as represented in leaf 2. Blood, epithelium, GI and heart also have the largest RE (Fig S3C in the Supplementary Materials). In general, our results are consistent with the literature indicating relevance of blood tissue in SLE, and further add genomic-level support to the relevance of other tissues concurrently with blood \cite{castellano2015local, fujio2020transcriptome, hoffman1980gastrointestinal, ebert2011gastrointestinal}.

\subsubsection{Cell-type-level investigation}
Based on the observed relationship between GS annotation for blood tissue and SLE, in the second phase of the real data analysis, we considered $10$ blood-related GSP functional annotations. With a GSP score cutoff of $0.5$, $25.29\%$ were annotated in at least one of the 10 GSP blood annotations (Fig S4A in the Supplementary Materials) and the highest enrichment was observed for primary regulatory T cells ($12.13\%$) (Fig S4B in the Supplementary Materials). The highest overlaps were observed between SNPs annotated with primary memory helper T, effector memory T and CD8$^+$ memory T cells (Fig S4C in the Supplementary Materials).

Next, we applied GPA-Tree to the SLE GWAS and GSP blood annotations. At the nominal global FDR level of 0.05, GPA-Tree identified $8,993$ SLE-associated SNPs, where $8,723$ among those overlapped with the SNPs prioritized in the first phase using GS annotations. Among the SLE-associated SNPs prioritized in the second phase, $37.54\%$ were annotated for at least one of the 10 GSP blood annotations (Fig S5A in the Supplementary Materials). The largest proportion of SLE-associated SNPs was annotated for primary B cells ($19.47\%$), followed by primary regulatory T cells ($18.45\%$) (Fig S5B in the Supplementary Materials). Primary B cells also showed the highest RE with the value of $2.12$ (Fig S5C in the Supplementary Materials). Since SLE is characterized by the production of autoantibodies, the involvement of B cells, which produce antibodies, is consistent with disease pathology.

The original GPA-Tree model with GSP blood annotations identified primary B cells at the root node and included 172 leaves. Again, to improve interpretability and visualization, we used ShinyGPATree to prune the tree so that it includes 10 leaf nodes (Fig \ref{fig:Figure 5}B). In addition to primary B cells, other blood-related GSP functional annotations identified as important included primary memory helper T, regulatory T, neutrophils, natural killer, effector memory T, and CD8$^+$ memory T cells. Among the $8,993$ SLE-associated SNPs, 613 are concurrently annotated for primary B and helper memory T cells as represented in leaf 8; 68 are concurrently annotated for primary B and CD8$^+$ memory T cells while not being annotated for memory helper T cells as represented in leaf 10; and 108 are concurrently annotated for primary regulatory T, neutrophils and effector memory T cells while not being annotated for primary B cells as represented in leaf 4. Overall, these results are consistent with previous literature indicating connections between SLE and B cells, regulatory T cells, neutrophils and CD8$^+$ memory T cells \cite{comte2015t, sanz2010b, kaplan2011neutrophils, blanco2005increase, filaci2001impairment}. 

These results also provide several new insights for future investigations. For instance, among the SLE-associated SNPs, $43$ SNPs located in the $CLEC16A$ gene and $41$ SNPs located in the $IKZF\: 3$ gene are in leaf 8 and concurrently regulate primary B and memory helper T cells; however, an additional 16 SNPs in the $CLEC16A$ gene are in leaf 4 and concurrently regulate primary regulatory T, neutrophils and effector memory T cells while not regulating B cells. These results provide further evidence that multiple independent SNPs in the same gene locus can have different effects on the levels of different immune cell subtypes \cite{ramos2020unravelling}, and can be utilized to investigate a variant's functional role in previously defined associations between SLE, $CLEC16A$ and $IKZF 3$ \cite{tam2015systemic, cui2013genetic, gateva2009large, lessard2012identification, manou2019235, cai2014association}, among others. 

\section{Conclusion}
In this paper, we presented GPA-Tree, a novel statistical methodology that integrates GWAS summary statistics and functional annotation data within a unified framework. GPA-Tree simultaneously identifies risk-associated SNPs and combinations of functional annotations that potentially explain the mechanisms through which risk-associated SNPs are related with traits. GPA-Tree showed the higher AUC and statistical power to detect risk-associated SNPs compared to existing approaches. 
GPA-Tree also successfully identified the true combinations of functional annotations in most cases, facilitating understanding of potential biological mechanisms linking risk-associated SNPs with complex traits. The proposed GPA-Tree approach was implemented as the R package \texttt{`GPATree'} and we also developed \texttt{`ShinyGPATree'}, a Shiny app for interactive and dynamic investigation of association mapping results and functional annotation trees. In the future, we plan to further improve the proposed GPA-Tree method to jointly analyze GWAS data for multiple traits \cite{kim2018improving, chung2017graph}. Overall, the ability of GPA-Tree to improve SNP prioritization and attribute functional characteristics to risk-associated SNPs or gene locus can be powerful in facilitating our understanding of genetic susceptibility factors related to complex traits. 

\section*{Funding}
This work was supported in part by NIH/NIGMS grant R01-GM122078, NIH/NCI grant R21-CA209848, NIH/NIDA grant U01-DA045300, and NIH/NIAMS grants P30-AR072582 and R01-AR071947.\\

\hspace{-0.4cm}\textit{Conflict of Interest:} None declared.
\vspace*{-12pt}

\bibliographystyle{plain}

\bibliography{main}

\newpage

\section*{Supplementary Material for ``GPA-Tree: Statistical Approach for Functional-Annotation-Tree-Guided Prioritization of GWAS Results"}
\setcounter{section}{0}
\section{Details of GPA-Tree}
Assuming the SNPs are independent, we can write the joint distribution of the observed data $Pr(\bf{Y}, \bf{A})$  as shown below. 
\begin{multline*}
\begin{array}{l@{}l}
P(\bf{Y}, \bf{A})
	&{}= \prod\limits_{i = 1}^{M} [ P(Z_i = 1) P(Y_i;Z_i = 1, A_{i.}) \: + (Z_i = 0) P(Y_i;Z_i=0, A_{i.}) ] \\
	&{} =\prod\limits_{i = 1}^{M} [ \pi_{i} \: \alpha \: y_{i}^{\alpha -1} + (1 - \pi_{i}) ]     
	\end{array}
\end{multline*}

The incomplete data log-likelihood ($\ell_{IC}$) and the complete data log-likelihood ($\ell_{C}$) for GPA-Tree are also shown below.
\begin{equation*}
\begin{array}{l@{}l}
	\ell_{IC} & {}= \sum\limits_{i = 1}^{M} log \left[ P(Z_i = 1) P(Y_i;Z_i=1) + P(Z_i = 0) P(Y_i;Z_i=0) \right]\\
	&{} = \sum\limits_{i = 1}^{M} log \left[ \pi_{i} \: \alpha \: y_{i}^{\alpha -1} + (1 - \pi_{i})  \right] 
	\end{array}
\end{equation*}

\begin{equation*}
\begin{array}{l@{}l}
\ell_{C}
&{}= \sum\limits_{i=1}^{M} \left[ Z_{i} \:( log \:\pi_{i} + log \:\alpha+ (\alpha-1) \: log \:y_{i} ) + (1-Z_{i})\: log(1-{\pi_{i}}) \right ] 
\end{array}
\end{equation*}

\section{Additional Figures}
\begin{figure}[!ht]
\includegraphics[width=\textwidth]{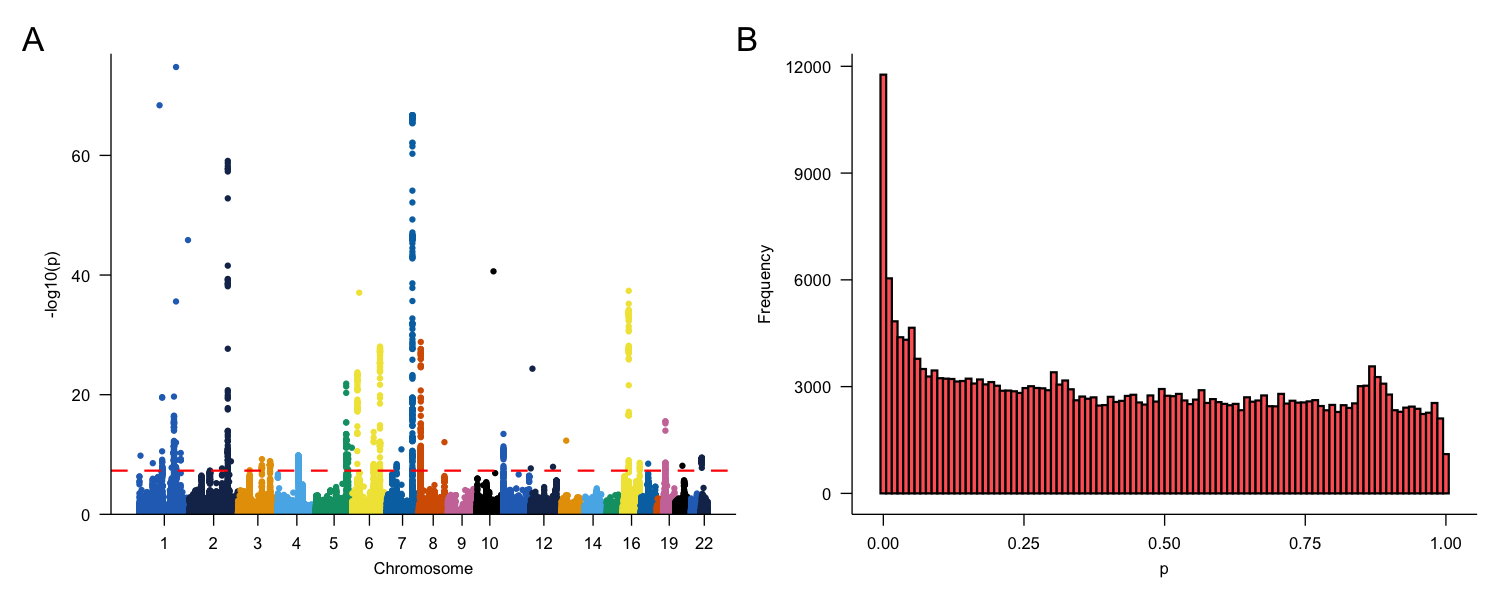}
\caption{Characteristics of the SLE GWAS data. (A) Manhattan plot. Genome-wide significance level ($5 \times 10^{-8}$) is indicated by the dashed red line. (B) GWAS association \textit{p}-value histogram.}
\label{fig:Figure 6}
\end{figure}

\begin{figure}[!ht]
\includegraphics[width=1\textwidth]{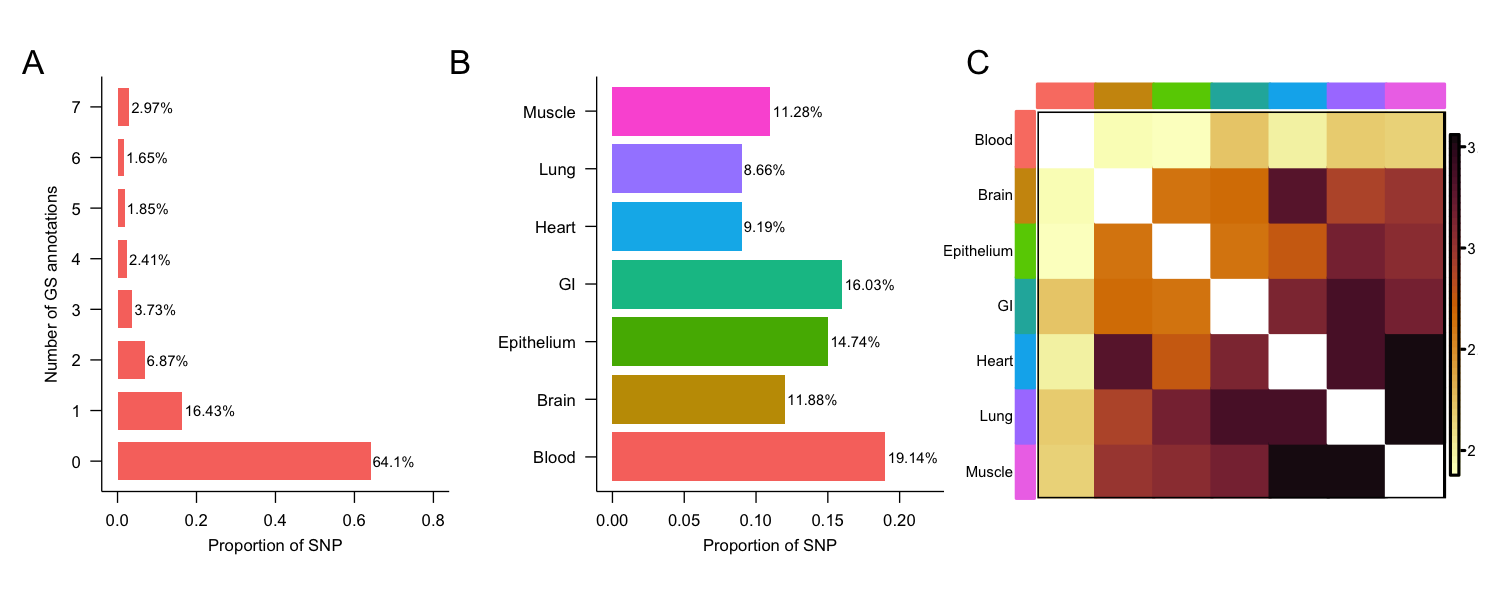}
\caption{Characteristics of $293,976$ SNPs when integrated with seven GenoSkyline (GS) annotations. (A) Number of GS tissues in which SNPs are annotated. (B) Proportion of SNPs that are annotated for each GS tissue type. (C) Overlap of SNPs annotated by seven GS tissue types, calculated using log odds ratio.}
\label{fig:Figure 7}
\end{figure}

\begin{figure}[!ht]
\includegraphics[width=1\textwidth]{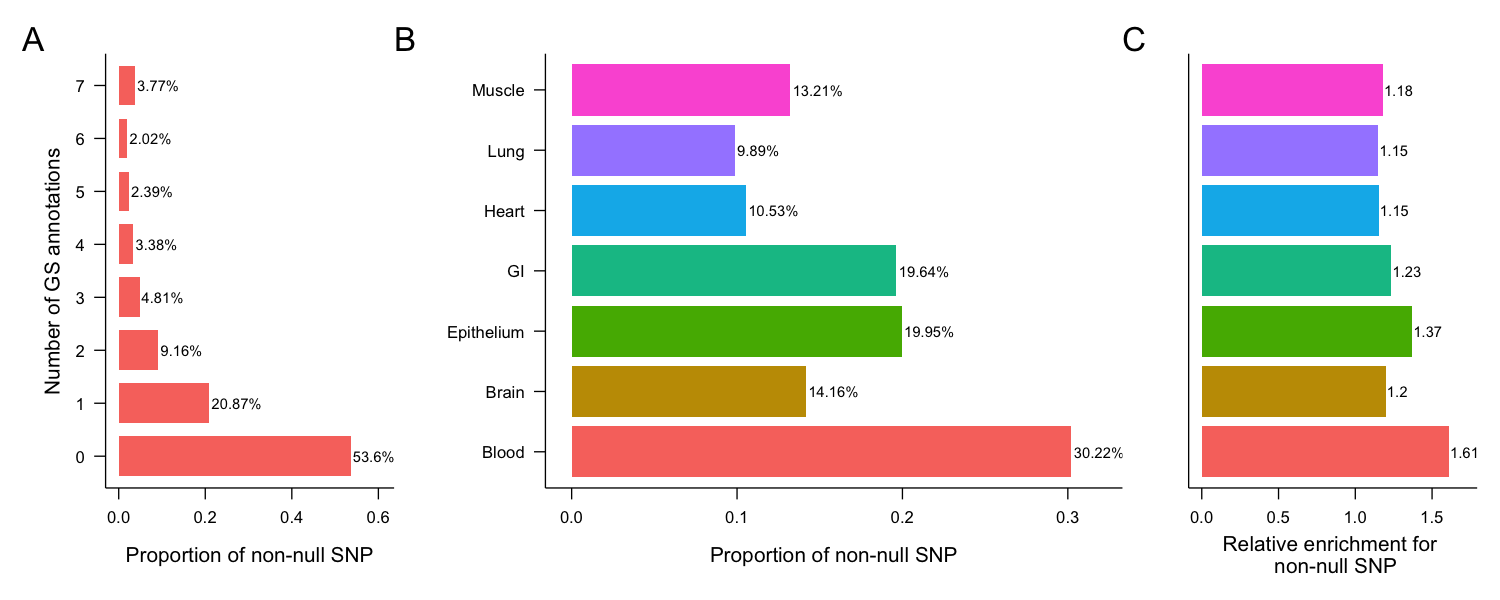}
\caption{Characteristics of the $8,962$ GPA-Tree identified SLE-associated SNPs when integrated with seven GenoSkyline (GS) annotations. (A) Number of GS tissues in which SLE-associated SNPs are annotated. (B) Proportion of SLE-associated SNPs annotated in each GS tissue type. (C) Relative enrichment (RE) of GS tissue types for SLE-associated SNPs. RE is defined as the ratio of the proportion of SLE-associated SNPs that are annotated for a specific GS tissue type, relative to the the proportion of non-SLE-associated SNPs that are annotated for the same GS tissue type.}
\label{fig:Figure 8}
\end{figure}

\begin{figure}[!ht]
\includegraphics[width=1\textwidth]{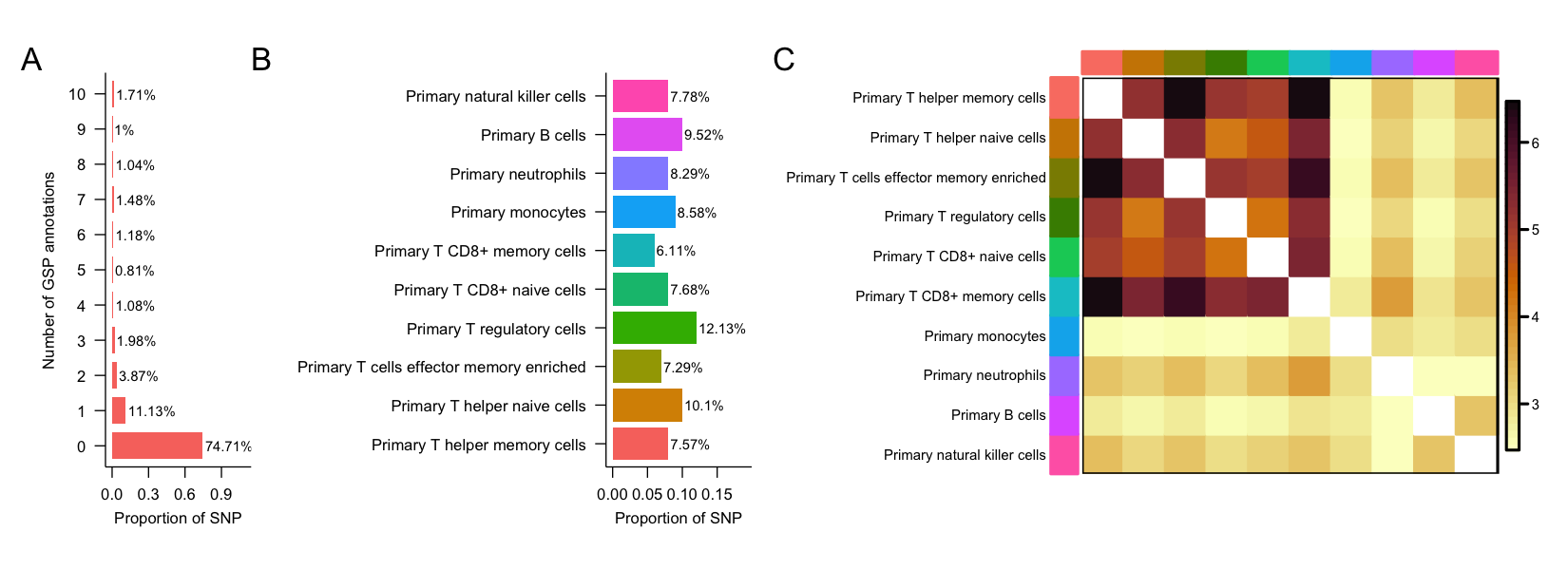}
\caption{Characteristics of the $293,976$ SNPs when integrated with 10 GenoSkylinePlus (GSP) blood-related annotations. (A) Number of blood-related GSP annotation type in which SNPs are annotated. (B) Proportion of SNPs annotated for each blood-related GSP annotation type. (C) Overlap of SNPs annotated by 10 blood-related GSP cell types, calculated using log odds ratio.}
\label{fig:Figure 10}
\end{figure}

\begin{figure}[!ht]
\includegraphics[width=1\textwidth]{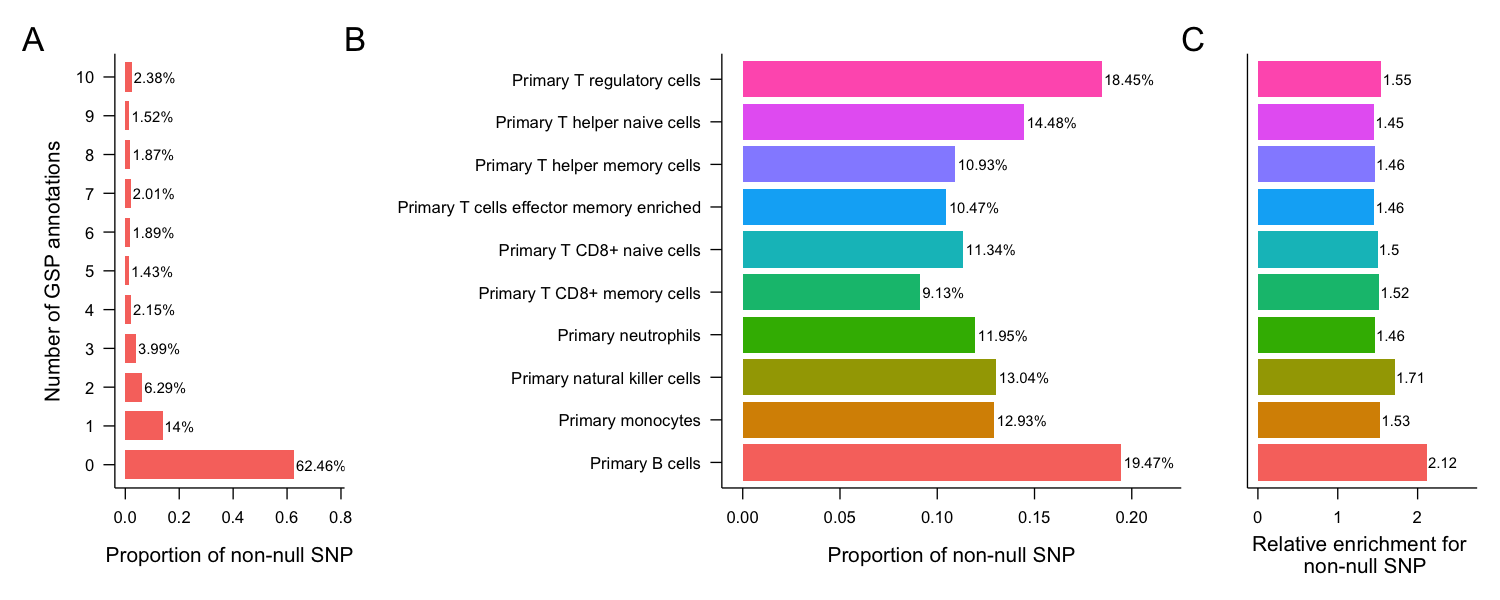}
\caption{Characteristics of the $8,993$ GPA-Tree identified SLE-associated SNPs when integrated with 10 blood-related GSP annotations. (A) Number of blood-related GSP annotations in which SLE-associated SNPs are annotated. (B) Proportion of SLE-associated SNPs annotated in each of the blood-related GSP annotation type. (C) Relative enrichment (RE) of blood-related GSP cell type for SLE-associated SNPs. RE is defined as the ratio of the proportion of SLE-associated SNPs that are annotated for a specific blood-related GSP cell type, relative to the the proportion of non-SLE-associated SNPs that are annotated for the same blood-related GSP cell type.}
\label{fig:Figure 11}
\end{figure}

\end{document}